\documentstyle[12pt]{article}
\begin{document}

\begin{titlepage}


\vspace{1cm}

\begin{center}
{\Large\bf Realistic  GUT with Gauge Mediated \\
Supersymmetry Breaking}

\vspace{3mm}

{\bf Zurab \ Tavartkiladze \footnote {E-mail address:
tavzur@axpfe1.fe.infn.it}
}
\vspace{3mm}

{\it INFN Sezione di Ferrara, I-44100 Ferrara, Italy \\
and Institute of Physics of Georgian Academy of Sciences,
380077 Tbilisi, Georgia
}
\end{center}

\begin{abstract}

We present an example of the gauge mediated SUSY breaking flipped
$SU(5)$ model. The messengers of the SUSY breaking are either only
colour triplets which belong to the minimal content of the scalar
supermultiplets or together with triplets as a messengers emerge the 
ordinary
Higgs doublets.
In both cases the model predicts light gauginos in respect of the
squarks and sleptons,
which could be tested in the
nearest LEP experiments.

In both cases ''all order`` solution of the doublet-triplet splitting
problem is obtained, the $\mu $-term of the order of $100$~GeV
is generated and the left handed neutrino masses are suppressed.

\end{abstract}

\end{titlepage}


The supersymmetric
theories  suggest
the elegant possibilities for solution of the gauge hierarchy problem.
Non-renormalization theorems \cite{wess} in SUSY theories imply
that certain ratios of coupling constants are non-renormalized in
exact SUSY limit.
This nice feature and  also the successful prediction of the
numerical value of $\sin^2 \theta_W$ \cite{lan} supported the idea of 
the
SUSY Grand Unified Theories (GUT).

The most interesting question is the origin of the SUSY breaking.
It is usually assumed that
SUSY is broken in a ''hidden`` sector and
by some interactions
transmitted in the visible sector. The most famous
scenarios are the supergravity theories \cite{barb} , in which the SUSY
breaking in the visible sector transmitted by the gravity. In this case 
the
soft SUSY breaking (SSB)
terms are presented at the
energies which correspond to the Planck scale - $M_P$ and
even they have the universal form,
they will renormalized
between the $M_P$ and $M_{GUT}$ .
At $M_{GUT}$ one has to integrate out the heavy particles and
evolve again the SSB parameters from $M_{GUT}$ to $m_W$ scale
with the RGEs of the MSSM. These processes violate the universalities
(see \cite{pom}
and references there)
and lead to the flavour changing neutral currents (FCNC).

To another class of the SUSY breaking scenarios belong the gauge 
mediated
supersymmetry breaking  (GMSB) models \cite{dine, fer} , in which the
supersymmetry breaking is transmitted by the gauge interactions.
Because the fact, that this models do not suffer from the FCNC problem
the interest in models of this type was  renewed recently
\cite{nel, amb} .

In this paper we present an example of the SUSY GUT in which the SUSY
breaking occurs in the sector of scalar superfields which are used
for the GUT symmetry breaking. 

The main contribution to the soft masses to the squarks and sleptons
comes from the nonzero $D$-term, which is just of the order of SUSY
scale; while gauginos gain masses through the
$SU(3)_C\times SU(2)_W\times U(1)_Y $ gauge 
interactions.

As a realistic model we consider the flipped $SU(5)$ theory which 
provides the
natural solution of the doublet-triplet (DT) splitting problem through 
the
missing partner mechanism . Crucial role in the SUSY breaking is played
by the anomalous $U(1)_A$ symmetry.

The model predicts gauginos with masses in the range of $1$~GeV, 
while the soft masses of the scalar particles are in the region
of $10^2-10^3$~GeV.

By the special implementation of the $U(1)_A$ charges of  some 
superfields the
model suggests two different sets of the messenger superfields.
In first case in the role of the messenger superfields
emerge the colour triplets and standard electroweak Higgs doublets.
which could give the nonuniversal contributions tu the masses
of the squarks and sleptons; However in the case considered
in the present paper these contributions are strongly suppressed.  
In this case the masses of all gauginos are generated through 
the one loop diagrams.
While in the second case in the role of the messengers we naturally have
only color triplets and no mass term generated for wino. 
In both cases the desirable $\mu $-term is generated, left handed 
neutrinos
are naturally light and proton decay through the $d=5$ operators
is strongly suppressed.

Let us note, that some examples in which the standard Higgs doublets
emerge as a messengers of the SUSY breaking also was considered
in the recent work \cite{mes} .

\vspace{0.5cm}

The flipped version of the $SU(5)$ model provides the solution
of the DT splitting problem through the missing partner mechanism by the 
most
economical way \cite{flip2, flip3} .

The gauge group is $SU(5)\times U(1)$ and the matter superfields 
transform
under this group as: $(10_1+\bar 5_{-3}+1_5)_i$ ($i$ is a family index )
in which the
ordinary quark and lepton superfields are compressed as:

$$
10_1=(q, d^c, \nu_R )_1~,
$$
$$
\bar 5_{-3}=(u^c, l)_{-3}~,
$$
\begin{equation}
1_5=e^c_5~.
\label{rep}
\end{equation}
The $10_1$ contains $\nu_R$ additional state  which is singlet under the
$ G_{321}\equiv SU(3)_C\times SU(2)_W \times U(1)_Y $ group.
The Higgs sector consists to the following superfields:

$$
H \sim 10_1=(Q, D^c, N)_1~,~~~~~~
\bar H \sim \overline {10}_{-1}=(\bar Q, \bar D^c, \bar N)_{-1}~,
$$
\begin{equation}
\phi \sim 5_{-2}=(T, h_d)_{-2}~,~~~~~~~
\overline {\phi } \sim \bar 5_{2}=(\bar T, h_u)_{2}~.
\label{screp}
\end{equation}
The $h_u$ and $h_d$ fields are generate the masses of up and down quarks
and leptons:

\begin{equation}
W^0_Y=10\cdot \bar 5 \cdot \overline {\phi } + 10\cdot 10 \cdot \phi
+1\cdot \bar 5 \cdot \phi ~.
\label{yuk1}
\end{equation}
the first term generates the masses of up type quarks, while the second 
and
third -  masses of down quarks and leptons respectively.

The $H+\bar H$  pair is  used for the GUT symmetry breaking. If $N+\bar 
N$
from  the set $H+\bar H$  develop  VEVs of order $M_X\simeq 10^{16}$~GeV,
then $SU(5)\times U(1)$
directly is broken to $G_{321} $ and
$Q(3, 2)+\bar Q (\bar 3 , 2)$ from $H+\bar H$ are eaten Goldstone modes.
The couplings between $H$ ($\bar H$) and $\phi $ ($\overline{\phi }$)
superfields are described
by the superpotential:

\begin{equation}
W_1^0=\lambda_1HH\phi+\lambda_2\bar H\bar H\overline {\phi }~.
\label{w1}
\end{equation}
Substituting the VEVs of the $N$ and $\bar N$ fields the mass terms
for the triplet components will get the form:

\begin{equation}
W_m=\lambda_1\langle N \rangle D^cT+
\lambda_2\langle \bar N \rangle \bar D^c\bar T~.
\label{wm}
\end{equation}
So, after the GUT symmetry breaking the triplet states decouple. While
$H$ and $\bar H$ do not contain doublet fragments,  $h_u$ and $h_d$
remain naturally light.

Suppose,  by  some mechanism (which will be presented below)
$N$ and $\bar N$ have F-terms with nonzero VEVs which
magnitudes are of the order $\sim mM_X$ ($m$
is mass scale up to which the SUSY is switched on). This will
cause the shift between the masses of the scalar and fermionic 
components
from the triplet superfields by the value $\sim \sqrt{mM_X}$.
While $mM_X\ll \langle N \rangle^2, \langle \bar N\rangle ^2$,
masses of the scalar components will
not changed (see (\ref {wm})). So, we will not have the light triplet 
states
in the spectra and the successful unification of the three gauge 
coupling
constants will not be altered. While the colour-triplet fragments
transform nontrivially under $SU(3)_C$ and $U(1)_Y$ gauge groups, the
SUSY breaking will transferred from the $W_1^0$ sector by the gauge
interactions.
The gauginos get masses through one loop diagrams. 
For general set of  messengers their masses are  given by the 
formula
\cite{nel}:

\begin{equation}
M_a=\frac{\alpha_a}{4\pi }\frac{F_i}{M}n_a(i)~,
\label{mgaug}
\end{equation}
where the $M$ is the mass of the corresponding messenger, and $F_i$
- appropriate $F$-term. $n_a(i)$ is Dynkin index and
for  fundamental representation of $SU(N)$ equals to 1 and  for
$U(1)_Y$, $n_1=\frac{6}{5}Y^2$ ($Y=Q_{em}-T_3$). 
Index $a$ in (\ref {mgaug}) corresponds to the gauge group and is $1$,
$2$ and $3$ for $U(1)_Y$, $SU(2)_W$ and $SU(3)_C$ respectively.
The (\ref {mgaug})
is written for $F_i \ll M^2$ case.

In our case only gluinos and $U(1)_Y$ gauginos get masses and winos
remain massless in the one loop level, since there is not doublets
among the messengers.

In this situation one state of chargino
is lighter then $W$ boson \footnote {This happens if the $\mu $ term 
exists
for the doublet components. If $\mu $-term is zero we will have two
light states in the theory. However as will be shown later in our model
the $\mu $ term with the desirable magnitude can be generated.}.
According to refs. \cite{far} - \cite{alon} this case did not excluded and 
requires
the low $\tan \beta $ regime .

The squarks and sleptons get masses through two loop diagrams \cite{dim}
and are given by the following formula:

\begin{equation}
\tilde{m}^2=2\left(\frac{F}{M} \right)^2
\sum \left(\frac{\alpha_a}{4\pi } \right)^2C_an_a~,
\label{msc}
\end{equation}
where $C_3=4/3$, $C_2=3/4$ and $C_1=3/5Y^2$.

\vspace{1cm}

Let us now describe how the SUSY breaking occurs in  our model.
For SUSY breaking we introduce the anomalous $U(1)_A$ gauge symmetry.
As a source of the SUSY breaking $U(1)_A$ symmetry was used in the 
recent
works \cite{ano}. The properties of the anomalous $U(1)_A$ symmetry also
were used for explaining the problem of gauge hierarchy  \cite{anDT, 
anDT1}
as well as for the understanding of
the pattern of fermion masses and mixing \cite{anDT1, ferm}.

It is well known, that anomalous extra $U(1)$ factors appear in 
effective
field theories from strings. The cancellation of its 
anomalies occurs by the
Green-Schwarz mechanism \cite{green} .
Because of the anomaly the Fayet-Iliopoulos term is
always generated \cite{fayet}
and in string theories it equals to \cite{wit}

\begin{equation}
\xi =\frac{g_A^2M_P^2}{192\pi^2}TrQ~.
\label{xi}
\end{equation}
So, $D_A$ term will have the form:

\begin{equation}
\frac{g_A^2}{8}D_A^2=\frac{g_A^2}{8}
\left(\Sigma q_i|\varphi_i |^2+\xi \right)^2~.
\end{equation}

In our model, which gauge group  is $G=SU(5)\times U(1)\times U(1)_A$,
the superfields have the following
prescription of the $U(1)_A$ charges:
$q_H=q_{\bar H}=-1$, $q_{\phi }=q_{\overline {\phi }}=2$.
This choice of the
charges will not change the form of the $W_1^0$ (see (\ref {w1})).
We also introduce the singlet superfields $X$,  $Y$ and $\bar Y$
with $U(1)_A$ charges: $q_X=2$ and $q_Y=q_{\bar Y}=0$.
The scalar superpotential

\begin{equation}
W_0=\lambda \frac{\bar Y Y}{M_P^2}X\bar HH
\label{w0}
\end{equation}
is the most general under $G\times {\cal R}$ symmetry, where ${\cal R}$
symmetry acts on superfields
$\varphi_i \to e^{{\rm i}R_{\varphi_i}}\varphi_i$
in such a way that $W\to e^{{\rm i}\alpha }W$.
So, the ${\cal R}$ `charges' of the superpotential
and superfields are arranged as follows:
$$
R_W=\alpha ~,~~~~R_{\overline{\phi }}=\alpha -2R_{\bar H}~,~~~~
R_{\phi }=\alpha -2R_H~,
$$
\begin{equation}
R_Y+R_{\bar Y}=\alpha -R_H-R_{\bar H }-R_X~,~~~~
\label{rs}
\end{equation}
so, $W_1^0+W_0$  (see (\ref {w1}) and (\ref {w0})) can
be the most general without
fixing $R$ numbers of the all superfields.
The potential builded from $F$ and $D$-terms will have the form:

\begin{equation}
V=\sum |F_i|^2+\tilde{g}^2\left(|H|^2-|\bar H|^2 \right)^2+
\frac{g^2_A}{8}\left(\xi - |H|^2-|\bar H|^2
+2|X|^2\right)^2~,
\label{v}
\end{equation}
where $\tilde{g}^2=\frac{3}{10}g^2+\frac{1}{8}g_1^2$ ($g$ and $g_1$
are the $SU(5)$ and $U(1)$ coupling constants respectively).
Supposing that $\xi >0$, one can easily
see that there exists the SUSY conserving
minima:

$$
\bar YY=0~,~~~|H|=|\bar H|~,~~~
$$
\begin{equation}
|H|^2+|\bar H|^2-2|X|^2=\xi ~;
\label{cl1}
\end{equation}
So, for the scalar superpotential given by (\ref{w0}) SUSY remains 
unbroken.

Let us imply the proposal of ref. \cite{ano} and suppose 
that the $\bar Y$
and $Y$ superfields transform nontrivially under the some
gauge group which interaction becomes strong below some 
$\Lambda$ scale .
The simplest case is the $SU(2)$ gauge group under which $Y$ and 
$\bar Y$
are the pair of doublet-antidoublet.
Non-perturbative superpotential induced by the instanton
effect have the form \cite{inst} :

\begin{equation}
W_{\rm inst}=\frac{\Lambda^5}{\bar Y Y}
\label{inst}
\end{equation}
and whole scalar superpotential  will be
\footnote{Non-perturbative term can violate the {\cal R} symmetry
if the {\cal R} symmetry is an anomalous.}
:

\begin{equation}
W=\lambda \frac{\bar YY}{M_P^2} X\bar H H
+ \frac{\Lambda^5}{\bar Y Y}~. \label{w} \end{equation}

The $F$ and $D$-terms will have the forms:

$$
F_H=\lambda \frac{\bar YY}{M_P^2}X\bar H ~,
~~~~F_{\bar H}=\lambda \frac{\bar YY}{M_P^2}XH ~,
~~~~F_X=\lambda \frac{\bar YY}{M_P^2}\bar HH~,
$$
$$
F_Y=\lambda \frac{\bar Y}{M_P^2}X\bar HH-
\frac{\Lambda^5}{\bar YY^2}~,~~~
F_{\bar Y}=\lambda \frac{Y}{M_P^2}X\bar HH
-\frac{\Lambda^5}{\bar Y^2Y}
$$
\begin{equation}
D=|H|^2-|\bar H|^2 ~,~~~
D_A=\xi - |H|^2-|\bar H|^2+2|X|^2~.
\label{fd}
\end{equation}
It is easy to see that SUSY is broken
because there is no solution with vanishing $F$ and $D$ terms.

Minimizing the potential, builded from the $F$ and $D$ terms
(see (\ref {fd} )), we can find that minimum can be obtained for
the solutions:

$$
H^2=\bar H^2 =\frac{3}{5}\xi ~,~~~
X^2=\frac{1}{10}\xi +\frac{5m^2}{g_A^2}~,
$$
\begin{equation}
\bar Y^4=Y^4=\frac{25}{3}M_P^4\frac{m^2}{\xi }
\left(1+\frac{125}{6\sqrt{3}\lambda }
\frac{mM_P^2}{\xi \sqrt{\xi }}\right)~.
\label{sols}
\end{equation}

where

\begin{equation}
m^2=\frac{2}{\sqrt{10}}\frac{\lambda \Lambda^5 }{M_P^2\sqrt {\xi }}
\label{pars}
\end{equation}
and for $\Lambda \sim 10^{11}-10^{12}$~GeV, $\sqrt {\xi }\sim 
10^{16}$~GeV
and $M_P \sim 10^{18}$~GeV we obtain $m \sim 100$~GeV-$10$~TeV.
For (\ref {sols})  solutions taking into account
(\ref {fd}), (\ref {pars})

$$
F_X\sim F_H=F_{\bar H}=m\sqrt {\xi }~,~~~
F_Y=F_{\bar Y}\sim
mM_P\left(\frac{m}{\sqrt {\xi }} \right)^{1/2}~,
$$
\begin{equation}
D=0~,~~~D_A=\frac{10m^2}{g_A^2}~.
\label{sol1}
\end{equation}

As we see the SUSY in broken and nonzero $F_H$, $F_{\bar H}$-terms are the
middle geometrical between the  $M_X$ (GUT scale) and $\sim 
m_W$ scales ;
Also the nonvanishing $D_A$-term
 with magnitude $m^2$
is generated and the main contribution in the masses of the
scalar components of the ordinary superfields comes from this
term 
\footnote {The model in which the soft masses for the matter particles
are generated from the nonvanishing part of the anomalous $D$-term
was considered in \cite {GA} .}
and equals to:

\begin{equation}
\tilde{m}^2_{\varphi_i }=\frac{5}{2}m^2q_i~,
\label{dmas}
\end{equation}
where $q_i$ is the anomalous $U(1)_A$ charge of the appropriate
$\varphi_i$ superfield. 
The non-universal contributions to the squark masses through the supergravity
corrections $\sim m^2\epsilon_X^2 $ (where 
$\epsilon_X =M_X/M_P$) for $\epsilon_X \sim 10^{-2}$ will be 
negligible.

The upper bound of the soft masses 
(which are also proportional to the $m^2$) of the electroweak Higgs
doublets could be obtained from the requirement of the electroweak
symmetry breaking  and related to the mass of the top quark.
Namely for $m_t=175-180$~GeV upper bound on $m^2_{H_u}$ is
 $\sim (350~{\rm GeV})^2$ \cite{bar}. For this order of $m^2$
the mass of the gluino and also 'Majorana' masses of wino and zino
are of the order of $1$~GeV. The recent analyses of percentage exclusion
of such a light gluino was presented in \cite{csik}, where results of 
\cite{far1} were performed. Existence of light (or massless)
wino leads to the one state of chargino lighter then $W$ boson. According
refs. \cite{far, alon} this case did not excluded and requires
the low $\tan \beta $ regime. As far as the light bino concerned its
phenomenological implications were described in refs. \cite{far} .

The GMSB example with light gauginos was presented in ref. \cite{moh}
and with light wino in ref. \cite{alon}; While in the recent \cite{raby}
works the models with a gluino as a lightest SUSY particle (LSP) were
considered.

While the gaugino masses are generated through the nonzero F-terms
(of the $N+\bar N$ components of the $H+\bar H$ pair) in the one loop
level, there magnitude will be $\alpha_a /(4\pi ) m$. 
So, the model predicts the 
gauginos with low soft masses in respect to the soft masses
of the squarks and sleptons.

\vspace{0.5cm}

From (\ref {dmas}) we see that the matter superfields 
must have the positive
$U(1)_A$ charges. This can easily obtained
if $W_1^0$ (see (\ref{w1})) will be rewritten to the form:

\begin{equation}
W_1=\lambda_1HH\phi+
\lambda_2 \frac{Z}{M_P}\bar H\bar H\overline {\phi }
\label{w11}
\end{equation}
and the Yukawa superpotential for ordinary quarks and leptons
will have the form:

\begin{equation}
W_Y=10\cdot \bar 5\cdot \overline{\phi }+
\frac{Z_1 }{M_P}10\cdot 10\cdot \phi +
\frac{Z_1 }{M_P}\bar 5\cdot 1\cdot \phi~,
\label{modyuk}
\end{equation}
where we have introduced $Z $ and $Z_1 $ superfields  with
anomalous $q_Z$  and $q_{Z_1 }$ charges
respectively and assumed that $Z$, $Z_1 $  share the
VEVs with $H$, $\bar H$ and $X$ fields in the $D_A$ term.
Therefore from (\ref {w11}) and (\ref {modyuk})

$$
q_{\bar 5}=-1+\frac{1}{2}q_{Z_1}+q_Z~,~~~~~
q_{10}=-1-\frac{1}{2}q_{Z_1}~,
$$
\begin{equation}
q_1=-1-\frac{3}{2}q_{Z_1}-q_Z~.
\label{muxts}
\end{equation}
For $1-q_{Z_1}/2 <q_Z<-1-3q_{Z_1}/2$ and $q_{Z_1 }<-2$ 
charges of the all matter
superfields will be positive.

Modification of the $W_1^0$ do not change the picture of the SUSY
breaking because the ratio of the `effective' $F$ term and the mass
of the messengers will be unchanged. As far as the masses of one pair
of triplet-antitriplet components are concerned, their magnitudes
are $\sim M_X^2/M_P $.
This threshold will not spoil the picture of unification and even
suggests the possibility of the obtaining small value of the 
$\alpha_s $.
Namely, for $\sin^2 \theta_W =0.2313 $ ,
$\alpha_s=0.11 $ is obtained
\footnote {I thank I. Gogoladze for bringing my attention to this issue. 
 },
which indeed coincides with the QCD sum analyses
\cite{shif}.

Until going to the fermion sector let us note that
taking into account   (\ref{w}), (\ref{sols})  and (\ref{w11}) 
the nucleon decay
parameter, which is the $(1,1)$ element of the inverse matrix of 
triplets-
$(\hat{M_T}^{-1})_{11}$ have the magnitude $\sim mM_P/M_X^3$
and therefore
nucleon decay through $d=5$ operator is strongly suppressed.

Turning to the fermion sector, we will see that $R$ charges can be
arranged in such a way that the $\mu $-term will have the desirable 
magnitude.

One of the nice feature of the flipped $SU(5)$ theory is that in
its framework,
because $e^c$ is identified as a singlet state of $SU(5)$, there do not
exists the dangerous relation $\hat{M_d}=\hat{M_e}$ which is 
concomitant to
the minimal $SU(5)$ theories. However, one can see that
$10\cdot \bar 5 \cdot \overline {\phi }$ coupling generate   the large
''Dirac`` mass for the neutrino. To suppress this mass the ''Majorana``
mass should be generated for
the $\nu_R $ state and then mass of the $\nu_L$ can be suppressed by the
universal seesow mechanism \cite{seesow} .

Here we introduce the additional fermionic states $\Psi \sim 10_1$,
$\overline{\Psi} \sim 10_{-1}$ and ${\cal N} \sim 1_0$ (for each 
generation).
Let us also introduce the three scalar superfields $S$, $S_1$
and $S_2$, which carry the
$U(1)_A$ charges. So they can also  contribute in the $D_A$-term  and
can share the VEVs together with $H$, $\bar H$,
$X$, $Z$ and $Z_1 $ states. So, if there do not
exist for them  couplings
in the superpotential the abovepresented picture of the SUSY
breaking will not changed. The transformation properties
under the $SU(5)\times U(1)\times U(1)_A\times {\cal R}$ symmetry
of all introduced superfields  are
presented in the Table 1.


\begin{table}
\caption{The superfield transformation properties }
\label{tab1}
$\begin{array}{|c|c|c|c|}
\hline
{\rm Fields} &SU(5)\times U(1) &U(1)_A &{\cal R} \\
\hline
&&&\\
H &10_1 &-1 & R_{\bar H}  \\
&&&\\
\bar H &\overline {10}_{-1} &-1 &    R_{\bar H}   \\
&&&\\
\phi &5_{-2} &2 & \alpha -2 R_H  \\
&&&\\
\overline {\phi } &\bar 5_2 &2-q_Z & \alpha -2R_{\bar H}-R_Z \\
&&&\\
X &1_0 &2 &  R_X \\
&&&\\

Y &1_0 &0 &    R_{Y} \\
&&&\\

\bar Y &1_0 &0 &  \alpha -R_H-R_{\bar H}-R_X-R_{Y} \\
&&&\\

Z &1_0 &q_Z &  R_Z \\
&&&\\

Z_1 &1_0 &q_{Z_1 } &  R_{Z_1 } \\
&&&\\

S &1_0 &q &R_S \\
&&&\\

S_1 &1_0 &q_1 &R_1 \\
&&&\\

S_2 &1_0 &2(q-q_1)-q_{Z_1 } &-\alpha +2(R_H+R_{\bar H}+
R_S+R_X-R_1)-R_{Z_1 } \\
&&&\\

10_i &10_1 &-1-\frac{1}{2}q_{Z_1 } & R_H-\frac{1}{2}R_{Z_1 } \\
&&&\\

\bar 5_i &\bar 5_{-3} &-1+\frac{1}{2}q_{Z_1 }+q_Z &
-R_H+2R_{\bar H}+R_Z+\frac{1}{2}R_{Z_1 } \\
&&&\\

1_i &1_5 &-1-\frac{3}{2}q_{Z_1 }-q_Z &
3R_H-2R_{\bar H}-R_Z-\frac{3}{2}R_{Z_1 } \\
&&&\\
\Psi & 10_1&1-\frac{1}{2}q_{Z_1 }+q-q_1  &
R_H+R_X+R_S-R_1-\frac{1}{2}R_{Z_1 } \\
&&&\\
\overline{\Psi } &10_{-1} &-1+\frac{1}{2}q_{Z_1 }-q
&\alpha -R_H-R_X-R_S+\frac{1}{2}R_{Z_1 } \\
&&&\\
{\cal N} &1_0 &q_1+\frac{1}{2}q_{Z_1 }-q &
\alpha  -R_H-R_{\bar H}-R_X-R_S-R_1+\frac{1}{2}R_{Z_1 } \\
&&&\\
\hline

\end{array}$
\end{table}


Under these assignments of charges
the Yukawa superpotential which generates masses of
the quarks and leptons and
also neutrinos ''Dirac`` and ''Majorana`` masses have the form:

\begin{equation}
{\cal W}_Y=W_Y+W_Y'~,
\label{bigyuk}
\end{equation}
where
\begin{equation}
W_Y'=A10\cdot \overline{\Psi }\cdot S\frac{X}{M_P}+
B\overline{\Psi }\cdot \Psi S_1 +C\Psi \cdot {\cal N}\cdot \bar H +
DS_2\cdot {\cal N}^2~.
\label{nu}
\end{equation}
$A,\cdots ,D$ are the Yukawa matrices, which 
elements can be assumed 
to be of the order of one.

The neutrino mass matrix will have the form:

\begin{equation}
\begin{array}{ccccc}
 & {\begin{array}{ccccc} \,\,\nu_L & \,\,~~ \nu_R & 
\,\,~~N_{\overline{\Psi}}&
\,\,~~ N_{\Psi } &\,\,~~ N_{{\cal N}}
\end{array}}\\ \vspace{2mm}
\hat{m}_{\nu }= \begin{array}{c}
\nu_L \\ \nu_R \\ N_{\overline{\Psi}}
\\N_{\Psi } \\ N_{{\cal N}} \end{array}\!\!\!\!\! 
&{\left(\begin{array}{ccccc}
\,\, 0 &\,\,~~h_u^0 &\,\,~~0  &\,\,~~0 &\,\,~~0 \\
\,\, h_u^0 &\,\,~~0 &\,\,~~\frac{XS}{M_P} &\,\,~~ 0 &\,\,~~ 0  \\
\,\,0&\,\,~~\frac{XS}{M_P} &\,\,~~0 &\,\,~~S_1 &\,\,~~0 \\
\,\,0&\,\,~~0 &\,\,~~S_1 &\,\,~~0 &\,\,~~\bar H \\
\,\,0&\,\,~~0 &\,\,~~0&\,\,~~\bar H & \,\,~~S_2 \end{array}\right)~.}
\end{array}  \!\!  ~~~~~
\label{matnu}
\end{equation}
This is the seesaw mass matrix which results the light neutrino
with a mass of order:

\begin{equation}
m_{\nu }\simeq M_P^2\frac{\langle h_u^0 \rangle ^2}{M_X^3}~.
\label{mnu}
\end{equation}

Suppressing the neutrino masses the superpotential (\ref{nu})
was used in which the $X$ superfield (with nonzero $F$-term)
has couplings with superfields which are transforming nontrivially
under the $G_{321}$ gauge group and could emerge as a messengers.
 
Some remarks about implications of this fact  should be done.

For simplicity let us consider the case of the one generation.
After integrating out the heavy states of $\overline{\Psi }+\Psi $, 
which masses are $\sim M_X$, the decoupled state of decuplet (let 
us denote it by $10_h$) will leave in $10$ by the weight 
$\frac{SX}{S_1M_P}\sim \epsilon_X $, while the light state 
(denoting it by $10_f$) contained in it approximately by the 
weight $1$. Taking into the account these facts the first two 
terms of the (\ref{nu}) can be rewritten as follows:

\begin{equation}
W_Y'^{(2)}=\frac{SX}{M_P}\left(\epsilon 10_h\cdot \overline{\Psi }
+10_f\cdot \overline{\Psi }\right)+
S_1\cdot \overline{\Psi }\cdot 10_h~.
\label{nu1}
\end{equation}
As we see the $\overline{\Psi }+10_h$ also emerge as a messengers but
there contribution to the gaugino masses are strongly suppressed:

\begin{equation}
\delta M_a \sim \frac{\alpha_a}{4\pi }m\epsilon_X^2~.
\label{dgau}
\end{equation}

Because there exists the matter coupling with messenger 
(see (\ref{nu1})) the soft masses for scalar components of $10_f$
arise in the one loop level and as was shown in \cite{see} is of the
order:

\begin{equation}
\delta m^2_{10} \sim \frac{1}{16\pi^2 }\epsilon_X^2
\frac{F^4}{M^6}~,
\label{dsq}
\end{equation}
which is also negligible in comparison of (\ref{dmas}).

Turning to the case of the three generations, without loosing the 
generality the last term of (\ref{nu1}) could be taken diagonal, while
the remaining terms in general are nondiagonal. As was shown in \cite{giu}
because this fact the sparticles gain soft masses through the tadpole
$D$-term of the $U(1)_Y$; However in our case this contribution 
is miserable:

\begin{equation}
\delta m^2_{\tilde{\phi_i}} \sim \frac{\alpha_1}{4\pi }
Y_{\tilde{\phi_i}}m^2\epsilon_X^4~.
\label{dsq1}
\end{equation}

\vspace{0.5cm}

We have not fixed yet the anomalous $q$ and $q_1$ as well as $R_S$
and $R_1$ charges
of the $S$ and $S_1$
superfields respectively. Two cases can be considered:

\vspace{0.5cm}

{\bf 1. $q=-5$ and $R_S=-\frac{5}{2}R_X$}

In this case the term which is permitted by the $U(1)_A$ and ${\cal R}$
symmetries is
\begin{equation}
W_{\mu }=\overline{\phi } \phi
\frac{X^3Z Z_1 S_1^2S_2}{M_P^7}~.
\label{mu1}
\end{equation}
Substituting VEVs
$\langle S_1 \rangle \sim \langle S_2\rangle \sim
\langle X \rangle \sim M_X$ and
taking into account that $M_X/M_P=\epsilon_X \sim 10^{-2}$ we will 
have
$\mu =M_X\epsilon_X^7 \sim 100$~GeV.
Also substituting the nonzero $F$-term of the $X$
superfield we will obtain the shift 
with the magnitude
$mM_X\epsilon_X^7$ between the masses
of the scalar and fermionic components from $\bar {\phi }$ and $\phi $
superfields , this means that the standard doublets
also emerge as a messengers and wino can get mass through the
one loop diagram \cite{nel}. In this case the masses of the gauginos
will be:

\begin{equation}
M_3=
4\frac{\alpha_3}{4\pi }\frac{F_H}{\langle N \rangle}
=\frac{\alpha_3}{\pi }m~,
\label{g11}
\end{equation}

\begin{equation}
M_2=
3\frac{\alpha_2}{4\pi }\frac{mM_X\epsilon_X^7}{\mu }
=3\frac{\alpha_2}{4\pi }m~,
\label{g12}
\end{equation}

\begin{equation}
M_1
=\frac{\alpha_1}{4\pi }\left(
4\frac{F_H}{\langle N \rangle}
\frac{4}{30}+3\frac{mM_X\epsilon_X^7}{\mu }\frac{9}{30}
\right)
=\frac{43}{30}\frac{\alpha_1}{4\pi }m~.
\label{g13}
\end{equation}
In (\ref{g11}) the factor $4$  emerges because there
are two pairs of the messengers 
and after
the substitution of the $F$-terms in (\ref{wm})
the combinator factor $2$ arise ; while
in (\ref{g12}) factor $3$ arise because in (\ref{mu1}) the field $X$
(with nonzero $F$-term) is in the third power (the same arguments were
taken into the account during the calculation of $M_1$).

So, without introducing the additional states of the messenger
superfields we can obtain the desirable pattern of SUSY breaking in the
framework of the flipped $SU(5)$ GUT, with successful DT splitting
and with $\mu $-term of the order of $100$~GeV. All messengers are from the
minimal content of the GUT supermultiplets.
Interestingly in this case the standard Higgs doublets
also belong to the  messenger superfields. Coupling of messenger doublets
with ordinary matter induce the nonuniversalities in the one loop level
but will be miserable in respect of (\ref{dmas})
\footnote{See also the general arguments of ref. \cite{mes}. }.

\vspace{0.5cm}

{\bf 2. $q=-\frac{4}{5}$ and $R_S=-\frac{2}{5}R_X$}

In this case the $\mu $-term generating coupling is:

\begin{equation}
W_{\mu }=\overline{\phi } \phi
\frac{S^3Z Z_1 S_1^2S_2}{M_P^7}~,
\label{mu2}
\end{equation}
which also gives $\mu \sim 100$~GeV.
In this case the doublets do not have couplings  with superfields
which $F$-terms have the nonzero VEVs.

Note, that in both cases the values of the $q_1$ and $q_Z$
charges still
were undetermined. In order to insure the nonzero VEVs for
$H$, $\bar H$ fields there
charges and the $\xi $ term in the $D_A$ (see (\ref {xi}), (\ref {fd}))
must have the different signs.
Therefore ${\rm Tr}Q =-46-28q_1+8q_Z-21q_{Z_1}/2 > 0$
and from this condition
we obtain $q_1 < (16q_Z-21q_{Z_1}-92)/56$.

Building our model we have assumed, that
$\epsilon_X=\sqrt{\xi}/M_P \sim 10^{-2}$
which for $M_P\sim 10^{18}$~GeV gives $\sqrt{\xi}\sim 10^{16}$~GeV, this 
value
was dictated from the scale of the grand unification. However,
for the flipped $SU(5)$ model derived from  strings \cite{flip2}
the preferable value
of $M_X$ is $10^{17}$~GeV  without loosing the
successful prediction of the $\sin^2 \theta_W $ \cite{unif}.
Increasing the value of $M_X$ the
picture of the abovepresented scenarios will not changed
if  for the values of $q$, $R_S$ for the two cases  $-13$,
$-13R_X/2$ and $-4/13$, $-2R_X/13$
will be taken
respectively. Then the couplings generating
the $\mu $-term
will be  $ X^{11}Z Z_1 S_1^2S_2/M_P^{15}\overline{\phi } \phi$
and
$ S^{11}Z Z_1 S_1^2S_2/M_P^{15}\overline{\phi } \phi$ respectively,
which for $\epsilon_X =\sqrt{\xi}/M_P \sim 10^{-1} $
still give $\mu \sim 100$~GeV.

\vspace{0.5cm}

{\bf {Acknowledgments}}

I am indebted to D. Comelli, G. Dvali and I. Gogoladze for 
discussions and important comments.

\end{document}